\documentclass[letters,usenatbib]{mn2e}
\usepackage{times}

\usepackage{natbib}
\usepackage[utf8x]{inputenc}
\usepackage{amsmath}
\usepackage{bm}
\usepackage{booktabs}
\usepackage{graphicx}

\newcommand{\refeq}[1]{Eq.~(\ref{#1})}
\newcommand{\refig }[1]{Fig.~\ref{#1}}
\newcommand{\reftab}[1]{Table~\ref{#1}}

\title[Dynamo action in thick disks around Kerr black holes]{Dynamo action in thick disks around Kerr black holes: high-order resistive GRMHD simulations}
\author[M. Bugli et al.]
{M. Bugli$^{1,2}$\thanks{matteo@mpa-garching.mpg.de}, L. Del Zanna$^{2,3,4}$, N. Bucciantini$^{3,4}$ \\
$^1$ Max-Planck-Institute f\"ur Astrophysik, Karl-Schwarzschild Strasse 1, 85741 Garching, Germany \\
$^2$ Dipartimento di Fisica e Astronomia, Università di Firenze, Via G. Sansone 1, 50019 Sesto F.no (Firenze), Italy \\
$^3$ INAF -- Osservatorio Astrofisico di Arcetri, L.go E. Fermi 5, 50125 Firenze, Italy\\
$^4$ INFN -- Sezione di Firenze, Via G. Sansone 1, 50019 Sesto F.no (Firenze), Italy }
\date{}

\begin{document}

\maketitle
\begin{abstract}
We present the first kinematic study of an \emph{$\alpha\Omega$-dynamo} in the \emph{General Relativistic Magneto-HydroDynamics} (GRMHD) regime, applied to thick disks orbiting around Kerr black holes and using a fully covariant mean field dynamo closure for the Ohm law. We show that the $\alpha\Omega$-dynamo mechanism leads to a continuous exponential growth of the magnetic field within the disk and to the formation of dynamo waves drifting away or toward the equatorial plane. Since the evolution of the magnetic field occurs qualitatively in the same fashion as in the Sun, we present also butterfly diagrams that characterize our models  and show the establishment of an additional timescale, which depends on the microscopic properties of the turbulent motions, possibly providing an alternative explanation to periodicities observed in many high-energy astrophysical sources where accretion onto a rotating black hole is believed to operate.
\end{abstract}
\begin{keywords}
magnetic fields -- MHD -- dynamo -- accretion -- relativistic processes -- plasmas.
\end{keywords}

\section{Introduction}
Ordered, large-scale magnetic fields are believed to be a fundamental ingredient of the accretion processes that power many of the astrophysical sources of high-energy emission, like jets from Active Galactic Nuclei \citep{McKinney:2009} or Gamma Ray Bursts \citep{Bucciantini:2009,Rezzolla:2011}. This is particularly true for the Blandford-Znajek mechanism \citep{Blandford:1977} where the rotational energy is extracted from a rotating black hole through large-scale magnetic fields penetrating the ergosphere and twisted by the rotation of the surrounding space-time. Unfortunately it is still not clear how such magnetic fields originate. The process of collapse to the compact objects which are believed to be the engines that power these sources could amplify any preexisting frozen-in field. However, recent MHD simulations of collapsing dense cores \citep{Hennebelle:2008,Santos-Lima:2012} have shown how the magnetic braking due to a preexisting large-scale magnetic field can extract angular momentum from the cloud fast enough to prevent the formation of a disk. Turbulent reconnection can efficiently inhibit the magnetic braking by removing magnetic flux, and therefore a turbulent magnetic field allows the formation of a disk. However, turbulent motions and small-scale instabilities in accretion disks, such as the \emph{Magneto-Rotational Instability} (MRI) \citep{Balbus:2003}, could enhance the level of turbulent magnetic fields, but the tangled final configuration would not be able to drive large scale outflows.

A solution for the origin of the necessary large-scale magnetic field is a \emph{dynamo} process. If one considers the presence of small-scale correlated fluctuations in the flow due to turbulence (that invariably arises in astrophysical plasmas with very high fluid and magnetic Reynolds numbers) a \emph{mean-field} dynamo mechanism \citep{Moffatt:1978} could lead to a large-scale effective electromotive force capable of generate and sustain a large-scale magnetic field. While this process has been vastly studied in classical MHD (mostly the solar dynamo),  there are only a few applications in general relativity. In particular, \citet{Khanna:1996} have tested the role of a differentially rotating absolute space in exciting dynamo modes in accretion disks in Kerr metric in the simplified regime where the displacement current is neglected and for a fixed velocity field. This approximation was shown to be reasonable by \citet{Brandenburg:1996}, but it cannot be achieved consistently in a covariant formalism, if causality is to be preserved. It might also not be appropriate when fast and variables motions, as expected in accretion, near the event horizon with velocities close to the speed of light are allowed to develop. The first formulation of a fully covariant closure of the GRMHD equations for a non-ideal plasma with a mean-field dynamo mechanism has been presented by \citet{Bucciantini:2013}, but so far it has never been applied to an actual astrophysical system. In this paper we present the first example (to our knowledge) of a resistive GRMHD simulation of a thick disk orbiting around a Kerr black hole with a mean-field dynamo mechanism. Given a small magnetic seed, we follow its evolution through the kinematic phase, assuming therefore that its feedback on the dynamics can still be neglected. We use the  ECHO code \citep{Del-Zanna:2007}, combined for the first time with a 3rd-order \emph{IMplicit-EXplicit} (IMEX) Runge-Kutta scheme \citep{Pareschi:2005}, to integrate the Maxwell equations in the \emph{3+1 formalism} \citep{Gourgoulhon:2012} using the fully covariant mean-field dynamo closure for the Ohm's law given by \citet{Bucciantini:2013}. 

The plan of this paper is as follow. In Section \ref{sec:dyn} we briefly describe the mean-field dynamo mechanism and its implementation in the ECHO code. Then in Section \ref{sec:simulation} we illustrate the setup of our simulations, and our choice for the dynamo properties. We finally show the results of our study and present our conclusions in section \ref{sec:conc}. In the following we set $c=G=1$ and absorb all the factors $\sqrt{4\pi}$ in the definition of the electromagnetic field.

\section{Dynamo model and equations}\label{sec:dyn}
For classical MHD, the introduction of a mean-field mechanism leads to the following form of the resistive-dynamo Ohm law:
\begin{equation}
\bm{E'}=\bm{E}+\bm{v}\times\bm{B}=\eta\bm{J}+\xi\bm{B},
\end{equation}
where $\bm{E}$, $\bm{B}$ and $\bm{J}$ are respectively the electric field, magnetic field and current measured by an eulerian observer, $\bm{E'}$ is the electric field measured in the frame comoving with the fluid-velocity $\bm{v}$, $\eta$ is the magnetic resistivity due mostly to mean-field effects (and in smaller part to collisions) and $\xi$ is the mean-field dynamo coefficient (most commonly employed in the literature as $\alpha_{\rm{dyn}}\equiv-\xi$). The latter term is a direct result of the assumption of correlated turbulent motions in the plasma, and provide the means for the generation of currents $\bm{J}'\equiv\xi/\eta\bm{B}$ parallel to the magnetic field $\bm{B}$, in contrast with the ideal case where $\bm{J}=(\bm{\nabla}\times\bm{B})\perp\bm{B}$. 
Following \citet{Bucciantini:2013}, we adopt a fully covariant form of Ohm's law for a resistive plasma with a mean-field $\alpha$-dynamo effect given by:
\begin{equation}\label{eq:ohm1}
e^{\mu}=\eta j^\mu+\xi b^\mu,
\end{equation} 
where $e^\mu$, $b^\mu$ and $j^\mu$ are respectively the electric field, magnetic field and current measured by an observer comoving with the fluid 4-velocity $u^\mu$. 
The spatial projection of \refeq{eq:ohm1} is then given by:
\begin{equation}\label{eq:ohm2}
\Gamma[\bm{E}+\bm{v}\times\bm{B}-(\bm{E}\cdot\bm{v})\bm{v}]=\eta(\bm{J}-q\bm{v}),
\end{equation}
and provides a mean to express the current $\bm{J}$ in terms of  $\bm{E}$, $\bm{B}$ and $\bm{v}$. Here $\Gamma\equiv(1-v^2)^{-1/2}$ is the Lorentz factor and $q=\bm{\nabla}\cdot\bm{E}$ is the local charge density. 

Since only the electric and magnetic fields evolve through time, we just need to integrate the Maxwell equations, which in the 3+1 formalism are:
\begin{equation}\label{eq:bmaxwell}
\gamma^{-1/2}\partial_{t}\left(\gamma^{1/2}\bm{B}\right)+
\bm{\nabla}\times(\alpha\bm{E}+\bm{\beta}\times\bm{B}) = 0,
\end{equation}	
\begin{equation}\label{eq:emaxwell}
\gamma^{-1/2}\partial_{t}\left(\gamma^{1/2}\bm{E}\right)+
\bm{\nabla}\times(-\alpha\bm{B}+\bm{\beta}\times\bm{E}) = -(\alpha\bm{J}-q\bm{\beta}),
\end{equation} 
with $\gamma$ the determinant of the spatial metric tensor $\gamma_{ij}$, $\alpha$ the lapse function and $\bm{\beta}$ the shift vector. Computing $\bm{J}$ from \refeq{eq:ohm2} and substituting it in \refeq{eq:emaxwell} we obtain the final form of the equation for the evolution of the electric field:
\begin{align}\label{eq:efield}
&\gamma^{-1/2}\partial_{t}\left(\gamma^{1/2}\bm{E}\right)=
\bm{\nabla}\!\times\!(\alpha\bm{B}\!-\!\bm{\beta}\!\times\!\bm{E})\!-\!(\alpha\bm{v}\!-\!\bm{\beta})(\bm{\nabla}\!\cdot\!\bm{E})\nonumber\\
& \!-\!\alpha\Gamma/\eta\{[\bm{E}\!+\!\bm{v}\!\times\!\bm{B}\!-\!(\bm{E}\!\cdot\!\bm{v})\bm{v}]\!-\!\xi[\bm{B}\!-\!\bm{v}\!\times\!\bm{E}\!-\!(\bm{B}\!\cdot\!\bm{v})\bm{v}]\}.
\end{align}

The above equation applies, rather than its ideal limit, when $\eta$ is not completely negligible, as it occurs when in addition to the usual ohmic diffusivity we take into account the turbulent contribution. In this case, we do expect the terms $\propto\eta^{-1}$ to influence heavily the evolution of the electric field, and since they  may evolve on a time-scale $\tau_{\eta}$ much shorter than the MHD time-scale some sort of implicit time integrator must be employed to guarantee stability to the integration of \refeq{eq:efield}. 

\begin{figure}
\centering
\includegraphics[scale=0.65, trim= 17 0 160 40, clip]{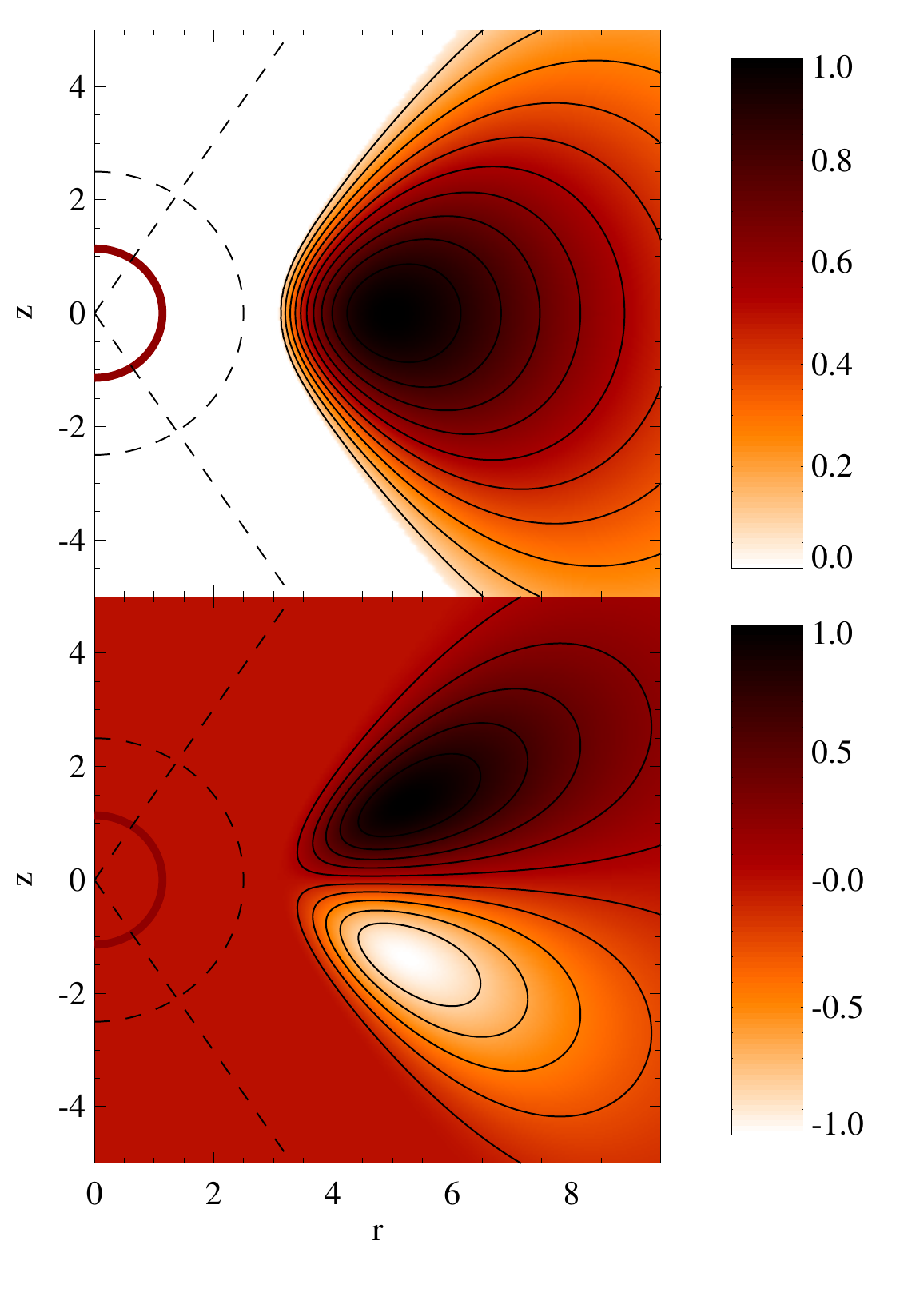}
\caption{Profiles and contours of the resistivity $\eta$ (upper panel) and dynamo coefficient $\xi$ (lower panel) normalized respectively to $\eta_{\rm disk} $and $\xi_{\rm disk}$. The red-solid line is the event horizon and the dashed lines represent the boundary of the computational domain.}\label{fig:disk_etaxi_profile}
\end{figure}

\section{Simulation Setup}\label{sec:simulation}
\subsection{The background model}
The assumption of kinematic dynamo implies that the background flow structure can be assumed as given. Our disk model corresponds to an unmagnetized thick torus with constant specific angular momentum $L$, in dynamical equilibrium, orbiting around a Kerr black hole (in Boyer-Lindquist coordinates) with specific angular momentum $a=0.99M_{\rm BH}$, as described in \citet{Font:2002}. Distances and time are expressed in units of $M_{\rm BH}$.

The center of the torus is located at a radius $r_{\rm c}=5$, while its inner edge is at $r_{\rm in}=3$. Here the temperature  is $p_{\rm c}/\rho_{\rm c}\simeq7\times10^{-3}$, where $p$ and $\rho$ are density and pressure. With these choices we determine $\rho,\bm{v}\text{ and } p$ at every point of the disk. The disk is assumed to be surrounded by a low density ($\rho_{\rm atm}\simeq 10^{-5}\rho_{\rm c}$ ) hot ($p_{\rm atm}/\rho_{\rm atm}\simeq0.5$)  atmosphere in hydrostatic equilibrium and corotating with the disk itself (so that the atmosphere is also in keplerian-like rotation). As we will show, the choice of the initial seed magnetic field is unimportant (apart from the parity with respect to the equator) for the evolution of the dynamo, given that the fastest growing mode will always be selected. Model can be initialized either  with a purely toroidal seed field ($B_T$) or a purely poloidal seed field ($B_P$). The initial field is always confined within the disk. The initial electric field is then set equal to the ideal value $\bm{E}=-\bm{v}\times\bm{B}$.

Given that in this letter we aim at presenting a first astrophysical application of a mean-field dynamo in GR to a  situation of interest, our choice for the dynamo properties is mostly dictated by simplicity.  We will therefore use the simplest possible choice for the parameter $\eta$ and $\xi$, that still preserve some of their expected global properties. We are aware that a sub-scale model for the turbulence will be required to build a proper realistic model, however, we hope to show here the feasibility of this approach, and derive a rough understanding of its working.

Because of the kinematic approximation we adopted, we expect the behavior of the $\alpha\Omega$-dynamo to be affected by the values of $\eta$ and $\xi$ alone, since the other free parameters of the model affect only fixed quantities. Here we investigate the response of the system to a variation of this two parameters. 

Let's consider the magnetic resistivity first. We assume the resistivity to be confined in the disk, where turbulent motions can act to dissipate the mean field. The resistivity in the disk scales as:
\begin{equation}
\eta(r,\theta)=\eta_{disk}\frac{\sqrt{\rho(r,\theta)}-\sqrt{\rho^*_{\rm atm}}}{\sqrt{\rho_{\rm c}}-\sqrt{\rho^*_{\rm atm}}},
\end{equation}
where the central density $\rho_{\rm c}$ is also the maximum density, $\rho^*_{\rm atm}$ is the minimum one in the atmosphere, and $\eta_{\rm disk}$ is the normalization value of the resistivity. The atmosphere is assumed to be an ideal conductor. The other option, corresponding to the opposed regime, is to assume it to be highly resistive, but we will not consider here this latter case. Since it is not possible to set $\eta=0$ using the IMEX schemes, we have set $\eta_{\rm atm}=10^{-5}$ (which is also a lower bound to the resistivity in the entire domain). We have verified that at the resolution of our runs, such value gives results indistinguishable from the ideal case (this is the value of our numerical resistivity).

In the same way we assume mean-field dynamo to act only inside the disk. The fact that the dynamo parameter $\xi$ never appears at the denominator in the resistive GRMHD equations allows us to set its value in the atmosphere to zero. Being $\xi$ proportional to the plasma kinetic helicity, it has therefore to be odd in the axial coordinate $z=r\cos\theta$, so that $\xi(r,\theta)=-\xi(r,\pi-\theta)$. 
In analogy with the resistivity, the simplest form we can choose is:
\begin{equation}
\xi(r,\theta):=
\begin{cases}
\xi_{\rm disk}\frac{\rho(r,\theta)\cos\theta}{(\rho\cos\theta)_{\rm max}}, & \text{inside the disk} \\
0, & \text{in the atmosphere}.
\end{cases}
\end{equation}
The profiles of resistivity and mean-field dynamo coefficient are shown in \refig{fig:disk_etaxi_profile}.

A mean-field dynamo process taking action in an accretion disk leads to a  $\alpha\Omega$-dynamo, where the toroidal magnetic field is enhanced by the differential rotation (the \emph{$\Omega$-effect}) and the poloidal component increases through the mean field dynamo (known in literature as the \emph{$\alpha$-effect}). In the kinematic case, the dynamics of such a system is characterized by the following \emph{dynamo numbers}:
\begin{equation}
C_{\xi}=\frac{\xi R}{\eta},\ \ \ \ \ C_\Omega=\frac{\Delta\Omega R^2}{\eta},
\end{equation}
which estimate the efficiency of the dynamo processes (due respectively to the mean-field dynamo and the differential rotation) in enhancing the magnetic field amplitude against the dissipation due to a finite resistivity of the plasma. In this relations $R$ represents a typical high-scale of the disk (for thick disks we used the radius of the center), while $\Delta\Omega$ is a typical shear (we have chosen the difference in angular velocity between the center and the inner edge of the disk).

\begin{figure}
\centering
\includegraphics[scale=0.57,trim=20 0 0 0,clip]{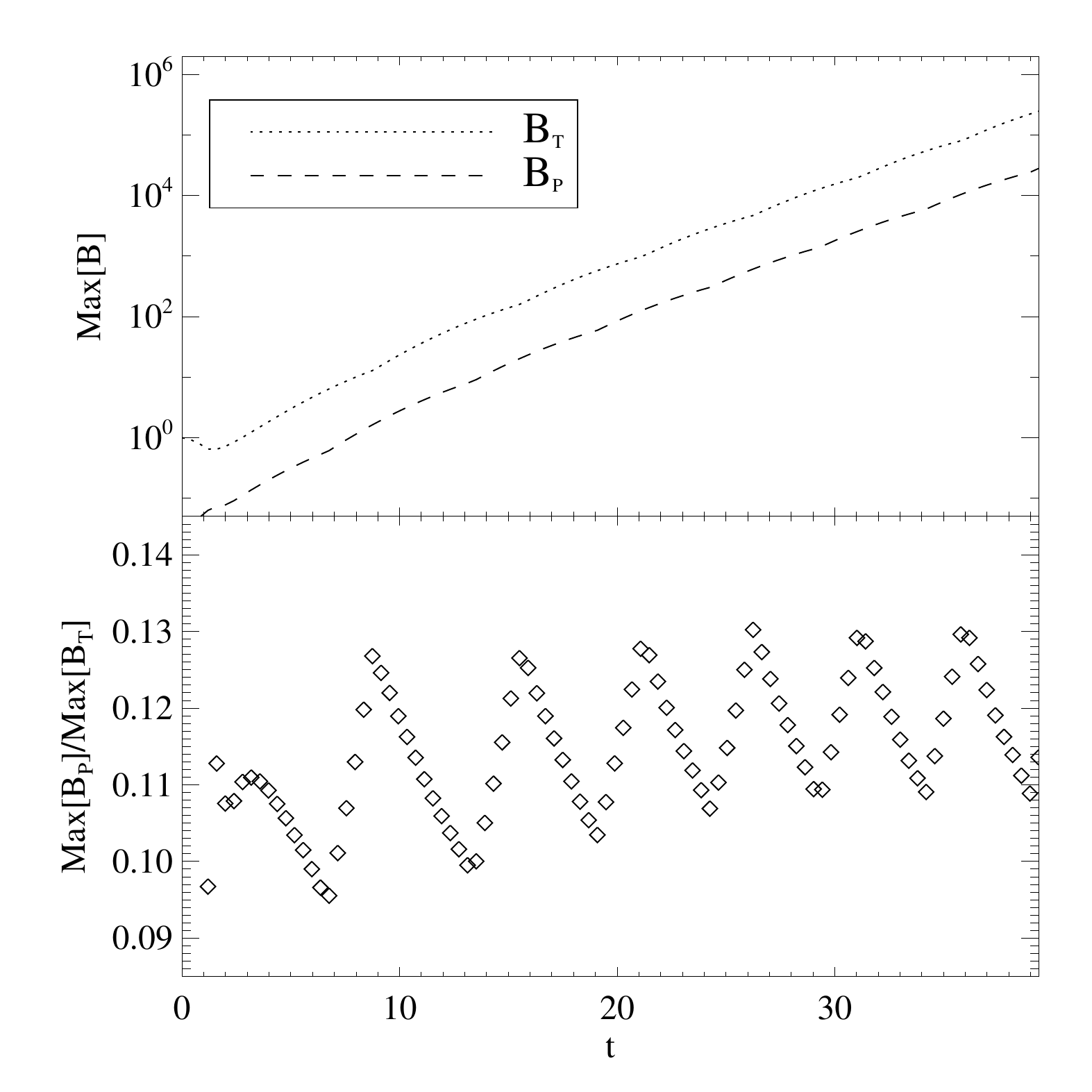}
\caption{Growth of $B_T=\sqrt{B^\phi B_\phi}$ and $B_P=\sqrt{B^r B_r + B^\theta B_\theta}$ (upper panel) and their ratio with time (lower panel) for Model 1. The magnetic fields are expressed in unit of the initial maximum value of $B_T$.The time is measured in units of the orbital period of the center of the disk $P_{\rm c}$.}\label{fig:bmax_bratio}
\end{figure}

\begin{figure*}\centering
\begin{minipage}{170 mm}
\centering
\includegraphics[scale=0.90, trim= 2 0 10 270, clip]{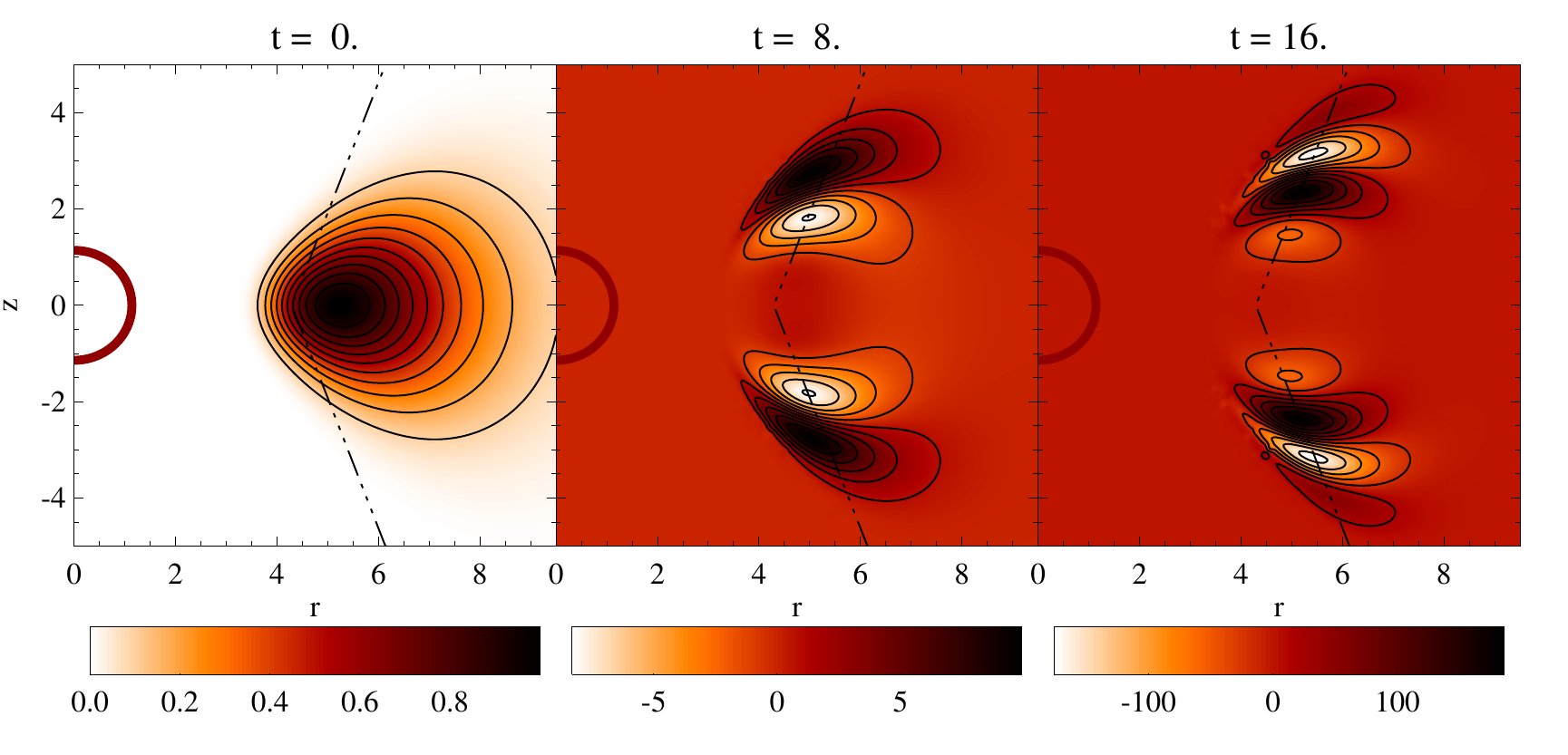}
\caption{Evolution of the toroidal component of the magnetic field for Model 1. The red-solid line is the event horizon, while the dashed lines represent the trajectories of the maxima of the magnetic field, inclined with respect to the equator by an angle $\pm\chi_0$. The magnetic field is measured in units of the maximum of the initial profile and the time in units of $P_{\rm c}.$}\label{fig:bphi}
\end{minipage}
\end{figure*}

\subsection{Numerical methods and settings}

Our domain extends in the range $r=[r_++1.5,25]$, $\theta=[\pi/4-0.2,3\pi/4+0.2]$, where $r_+$ is the radius of the event horizon. All of our simulations were performed on a numerical grid of $256\times256$ points. The grid is logarithmically stretched in the radial direction in order to have a larger resolution near the black hole (222 grid points within $r=10$). The domain has been shaped to fully contain the disk. We have verified that the magnetic field never migrates beyond the polar regions that we have excised. We have also verified that at this resolution the results are converged.

Developing what has been done by \citet{Bucciantini:2013}, we adopted an IMEX Runge-Kutta scheme \citep{Pareschi:2005,Palenzuela:2009} to perform the time integration of \refeq{eq:bmaxwell} and \refeq{eq:emaxwell}. In particular, extending previous results,  we have implemented the \emph{SSP3(4,3,3)} scheme, that guarantees third order accuracy in time. This has been coupled, for the hyperbolic part, to a HLL solver with a $5^{th}$ order spatial reconstruction routine (MP5), which allow us to reach a global $3^{rd}$ order accuracy and to get convergent results already at modest resolutions (for more comprehensive numerical tests see \citealp{Del-Zanna:2014}).

\section{Results and discussion}\label{sec:conc}
In \reftab{tab:results} we show the various runs of our study. They differ by the value of $\eta$, $\xi$ and the initial profile of the magnetic field. In all of the models where the initial magnetic field is toroidal (symmetric to the equator), the $\alpha$-effect generates from the beginning a poloidal antisymmetric component: therefore the magnetic field has quadrupolar symmetry. On the other hand, starting with a poloidal field (antisymmetric to the equator) a toroidal antisymmetric component arises due to the $\Omega$-effect, thus the field has dipolar symmetry. 

In either cases the magnetic field evolves to an eigenstate of the system, that is reached after a relaxation-time that ranges, depending on the particular model, roughly between  $3\ P_{\rm c}$ and $30\ P_{\rm c}$, with $P_c\simeq76.2 GM_{BH}/c^3$ the orbital period of the disk center. As shown in \refig{fig:bphi}, the eigenstate is characterized by a \emph{dynamo wave} that propagates away from the equatorial plane for $\xi>0$. For the models with $\xi<0$ the propagation occurs towards the equatorial plane, qualitatively in agreement with the solar dynamo case and consistent with our definition of $\xi\equiv-\alpha_{\rm{dyn}}$ \citep{Moffatt:1978}. The magnetic field appears to originate approximately where the dynamo parameter is stronger, and then drifts from that location. As the field migrates toward the atmosphere, or toward the equator (depending on the sign of $\xi$), resistive effects become proportionally stronger and the field is dissipated.

Besides this drifting, the amplitude on the poloidal and toroidal components of the magnetic field  grow exponentially with the same growth rate. This because, in the kinematic regime we have assumed, there is no quenching effect. The growth rate increases for larger values of $\xi$. The ratio between the maximum of the poloidal component and the maximum of the toroidal one ranges from $0.04$ to $0.36$ and shows an oscillating behavior indicative of a phase difference between these two components.

\begin{figure*}\centering
\begin{minipage}{170 mm}
\centering
\includegraphics[scale=0.95, trim= 3 0 10 310, clip]{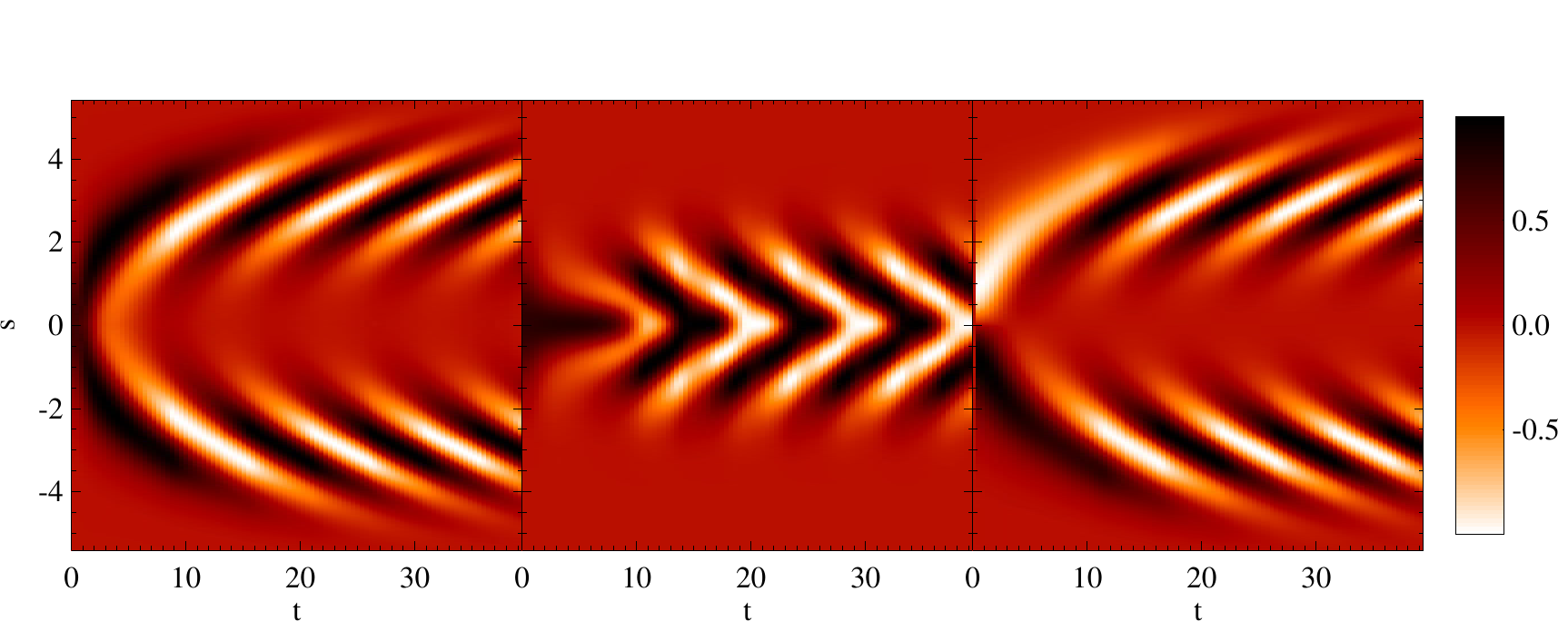}
\caption{Butterfly diagrams for the Models 1 (left), 2 (center) and 7 (right), showing the value of the toroidal magnetic field along the trajectories of the drifting (shown in \refig{fig:bphi}) as a function of time. The magnetic field has been normalized to its maximum value in the whole domain at every time, which is measured in units of the orbital period of the center of the disk $P_{\rm c}$. The parameter $s\equiv z/\sin\chi_0$ represents the position on the trajectories.}\label{fig:butterfly}
\end{minipage}
\end{figure*}

In order to quantitatively characterize the migration of the magnetic field within the disk, we produced for each model a \emph{butterfly diagram} considering the value of the toroidal field along the trajectories of its maxima at different times. From the diagrams for Models 1, 2 and 7, shown in \refig{fig:butterfly}, is now evident the periodicity of the eigenstate and its spatial scale. For larger values of $\xi$ we retrieve smaller periods and smaller spatial scales, while decreasing the action of the mean field dynamo both periods and spatial scales increase. 

\begin{table*}\centering
\begin{minipage}{140 mm}
\caption{Values of the parameters and the dynamical characteristics of each Model.}\label{tab:results}
\label{tab:results}	\centering
\begin{tabular}{ccccccccccc}
\toprule
& $\bm{B}_{\textrm{init}}$ & $\eta_{disk}$ & $\xi_{disk}$ & $C_\xi$ & $C_\Omega$ & Growth Rate & Period & $B_P/B_T$ & $s_{max}$ & $\chi_0$\\ 
\toprule
Model 1 & $B_T$ & $10^{-3}$ & $10^{-3}$ & 5 & 400 & 0.30 & 10.34 & 0.12 & 2.90 & 1.22 \\ 
Model 2 & $B_T$ & $10^{-3}$ & $-10^{-3}$ & -5 & 400 & 0.25 & \hphantom{0}9.14      & 0.12 & 0.67 & 1.41\\ 
Model 3 & $B_T$ & $10^{-3}$ & $5\times10^{-3}$ & 25 & 400 & 1.26 & \hphantom{0}3.31 & 0.36 & 2.42 & 1.14 \\ 
Model 4 & $B_T$ & $10^{-3}$ &   $2\times10^{-4}$ & 1 & 400 & 0.06 & 33.14 & 0.04 & 3.18 & 1.26\\ 
Model 5 & $B_T$ & $5\times10^{-3}$ & $5\times10^{-3}$ & 5 & 80 & 0.37 & \hphantom{0}6.63 & 0.17 & 3.11 & 1.26\\ 
Model 6 & $B_T$ & $2\times10^{-4}$ & $2\times10^{-4}$ & 5 & 2000 & 0.21 & 17.23      & 0.07 & 2.64 & 1.22\\ 
Model 7 & $B_P$ & $10^{-3}$ &   $10^{-3}$ & 5 & 400 & 0.30 & 10.34 & 0.12 & 2.95 & 1.18 \\ 
Model 8 & $B_P$ & $10^{-3}$ & $5\times10^{-3}$ & 25 & 400 & 1.26 & \hphantom{0}3.31 & 0.36 & 2.36 & 1.14\\ 
Model 9 & $B_P$ & $10^{-3}$ & $2\times10^{-4}$ & 1 & 400 & 0.06 & 34.47 & 0.04 & 3.13 & 1.26\\ 
Model 10 & $B_P$ & $10^{-3}$ & $-10^{-3}$ & -5 & 400 & 0.25 & \hphantom{0}9.54 & 0.12 & 0.82  & 1.37 \\ 
\bottomrule
\end{tabular}
\end{minipage}
\end{table*}

It is interesting to note that the particular symmetry of the initial magnetic fields does not affect the dynamical characteristics of the eigenstate selected by the problem: comparing for example the results for Models 1 and 7 (which differ only by the initial field) we can see how the growth rate, the period, the spatial scale and the ratio between poloidal and toroidal component are roughly the same. Moreover, observing the butterfly diagrams is evident how the initial symmetry of the magnetic field at the passage at the equator (\emph{quadrupolar} for the first six models, \emph{dipolar} for the rest of them) is preserved during all the evolution of the system. This means that the state reached by the system is degenerate in the parity at the equator of the magnetic field.

The turbulence proprieties, that determine the characteristics of the mean-field dynamo, can in principle set a time-scale (the dynamo wave period) which might be unrelated to the large-scale dynamics. Of course the same turbulence that sets the dynamo coefficient $\xi$, as well as the mean field resistivity $\eta$, should be related with other properties of the disk like the typical viscosity, and might be influenced by the cooling properties and the ionization state of the plasma. Understanding these possible connections, and their observable consequences, is however beyond the scope of this paper. We hope to extend this initial introductory work to a more realistic regime, where a more meaningful choice for $\eta$ and $\xi$, based on a turbulent model, could allow us to properly evaluate the importance of mean-field effects \citep{Brandenburg:2002}. This could help our understanding of some of the unexplained time-variability observed in accreting systems \citep{Gilfanov:2010}, that cannot be naively associated with orbital periodicities. 

The relation between mean-field dynamo and MRI in thick disks is also a fundamental aspect to be investigated. So far, this has been studied numerically only in the classical MHD limit in the \emph{shearing box} local approximation \citep[for a review][]{Blaes:2013}. When vertical stratification is included, magnetic buoyancy couples with the turbulent dynamo effects \citep{Brandenburg:1995} and quasi-periodical field amplification, migration, and reversal can be observed \citep{Davis:2010,Flock:2012}. It would be interesting to reproduce these effects of self-generated turbulence in the GRMHD case as well.

Further developments might also include:  the quenching of the $\alpha$ effect; 
the dynamical feedback of the magnetic field on the evolution of the hydrodynamical quantities, requiring the integration of the full set of GRMHD equations. Moreover, in this study we have fixed not only the disk structure but also the mass and the spin of the black hole. The role of these parameters needs also to be investigated, to understand if there are possible observable quantities that could help us to constrain them.  This could be of potential interest especially in view of the large improvements in precision of the measurements of  the mass and the spin of a black hole of the last few years \citep{Risaliti:2013}.

\bibliographystyle{mn2e}
\bibliography{bugli13}

\end{document}